\documentclass{aastex631}

\shorttitle{AASTeX v6.3.1 Sample article}
\shortauthors{Guest et al.}
\graphicspath{{./}{figures/}}

\begin{document}

\title{An X-ray Proper Motion Study of the LMC SNR 0509-67.5}

\correspondingauthor{Benson Guest}
\email{bguest1@umd.edu}

\author[0000-0003-4078-0251]{Benson T. Guest}
\affiliation{Department of Astronomy, University of Maryland, College Park, MD 20742, USA}
\affiliation{NASA Goddard Spaceflight Center, Greenbelt, MD 20771, USA}
\affiliation{Center for Research and Exploration in Space Science and Technology, NASA/GSFC, Greenbelt, MD 20771, USA}

\author[0000-0002-2614-1106]{Kazimierz J. Borkowski}
\affiliation{North Carolina State University, Raleigh, NC 27607, USA}

\author[0000-0002-9886-0839]{Parviz Ghavamian}
\affiliation{ Department of Physics, Astronomy and Geosciences, Towson University, Towson, MD 21252, USA}

\author[0000-0003-3850-2041]{Robert Petre}
\affiliation{NASA Goddard Spaceflight Center, Greenbelt, MD 20771, USA}

\author[0000-0002-5365-5444]{Stephen P. Reynolds}
\affiliation{Department of Physics, North Carolina State University, Raleigh, NC 27695, USA}

\author[0000-0002-5044-2988]{Ivo R. Seitenzahl}
\affiliation{School of Science, University of New South Wales, Australian Defence Force Academy, Canberra, ACT 2600, Australia}

\author[0000-0003-2063-381X]{Brian J. Williams}
\affiliation{NASA Goddard Spaceflight Center, Greenbelt, MD 20771, USA}


\begin{abstract}
We present a third epoch of \textit{Chandra} observations of the Type Ia Large Magellanic Cloud Supernova remnant (SNR) 0509-67.5. With these new observations from 2020, the baseline for proper motion measurements of the expansion has grown to 20 years (from the earliest {\it Chandra} observations in 2000). We report here the results of these new expansion measurements. The lack of nearby bright point sources renders absolute image alignment difficult. However, we are able to measure the average expansion of the diameter of the remnant along several projection directions. We find that the remnant is expanding with an average velocity of 6120 (4900 -- 7360) km s$^{-1}$. This high shock velocity is consistent with previous works, and also consistent with the inference that 0509-67.5 is expanding into a very low density surrounding medium. At the distance of the LMC, this velocity corresponds to an undecelerated age of 600 yrs, with the real age somewhat smaller.

\end{abstract}



\section{Introduction} \label{sec:intro}
The supernova remnant (SNR) 0509-67.5 was discovered in the Large Magellanic Cloud (LMC) by an X-ray survey using the Einstein observatory (\cite{Long1981}). Followup optical observations (\cite{Tuohy1982}) identified the remnant as the result of a Type Ia supernova. This was later supported by the \textit{Advanced Satellite for Cosmology and Astrophysics (ASCA)} observations in X-ray (\cite{Hughes1995}). The first \textit{Chandra} observations were presented in \cite{Warren2004} who found a circular morphology with the bulk of the continuum emission coming from non-thermal origins. They support the earlier conclusion from \cite{Tuohy1982} that the remnant is expanding into a low density environment. Further support for a low density environment surrounding 0509-67.5 was provided by {\it Spitzer} infrared observations, where modeling of emission from warm dust grains in the post-shock environment imply a pre-shock density of less than 1 cm$^{-3}$ \citep{Borkowski2006,Williams2011}.

A light echo was found by \cite{Rest2005} from which they estimate an age of 400 $\pm 120$ yr. \cite{Ghavamian2007} used spectroscopic UV observations to detect significant broadening of the Ly$\beta$ line, corresponding to a shock velocity of 5200-6300 km s$^{-1}$. An optical proper motion study using \textit{HST} H$\alpha$ observations over a 1 year baseline was presented by \cite{Hovey2015}. They find a global shock speed of 6500 $\pm$ 200 km s$^{-1}$. Their accompanying hydrodynamic simulations predict an age of 310 $^{+40}_{-30}$yr, consistent with that of \cite{Rest2005}, and in agreement with the calculations of \cite{Seitenzahl2019} which required an age of 310 years and a sub-Chandrasekhar explosion mass in order to simultaneously match the observed broad optical coronal line emission of Fe XIV and Fe XV, and the position and speed of the forward shock. \cite{Roper2018} used {\it Chandra} observations over a 7 year baseline (2000 and 2007) to perform an X-ray proper motion study. They report an overall average expansion velocity of $7500 \pm 1700$ km s$^{-1}$.

In this paper, we report results from a proper motion study using a new epoch of \textit{Chandra} X-ray observations from 2020. These observations extend the baseline of measurements to 20 years following the initial observations in 2000. The paper is organized as follows. In Section~\ref{observations}, we detail the X-ray observations and data reduction, the unsuccessful attempts to align the two epochs to a common coordinate system, and the alternative methods we employed. We report our results in Section~\ref{measurements}.

\section{Observations}
\label{observations}
A new epoch of \textit{Chandra} observations (PI: B. Williams) was acquired between April 11 and November 21 2020 totalling 414ks and spread over 14 individual segments ranging in length from 25-40ks (Table \ref{tab:obs}). The initial Chandra observations from May 12 2000 (PI: J. Hughes) consist of a single 49ks pointing. Both observations place SNR 0509-67.5 on the S3 chip of the ACIS-S array close to the optical axis of the telescope. 

\begin{table}[]
    \centering
    \begin{tabular}{c c c}
       obsID  & year & Exp (ks) \\\hline
\textbf{ 776}	&	\textbf{2000}	&	\textbf{48.99}	\\\hline
7635	&	2007	&	32.74	\\
8554	&	2007	&	29.47	\\\hline
\textbf{22442}	&	\textbf{2020}	&	\textbf{34.6}	\\
\textbf{22443}	&	\textbf{2020}	&	\textbf{36.58}	\\
22444	&	2020	&	24.74	\\
23020	&	2020	&	24.77	\\
23021	&	2020	&	27.7	\\
23022	&	2020	&	24.74	\\
23023	&	2020	&	24.59	\\
23024	&	2020	&	24.47	\\
\textbf{24635}	&	\textbf{2020}	&	\textbf{32.48}	\\
\textbf{24636}	&	\textbf{2020}	&	\textbf{32.17}	\\
\textbf{24637}	&	\textbf{2020}	&	\textbf{39.54}	\\
24638	&	2020	&	24.8	\\
\textbf{24858}	&	\textbf{2020}	&	\textbf{35.6}	\\
24867	&	2020	&	27.17	\\ \hline\hline
    \end{tabular}
    \caption{Available \textit{Chandra} observations. Those in bold were used in this work.}
    \label{tab:obs}
\end{table}

\subsection{Image Alignment and Reprojection}
The standard procedure for aligning observations requires the matching of known sources from external catalogues with point sources detected in the image, or aligning common point sources within images from each epoch.
We used the CIAO tool wavdetect to search for point sources in the reprocessed event files. The found point sources were then filtered by eye to exclude false positives and a transformation matrix file was created using the tool wcs\_match, and wcs\_update to align the 2000 observation and all 2020 segments to the longest observation from the new epoch(ObsID 24637).  This method was used in the analysis of \cite{Roper2018} using the 2000 and 2007 data. However, we find that the results of this alignment were not accurate for a robust measurement at the sub-pixel level. The point source brightnesses in the individual segments did not yield enough counts to achieve convincing fits to the point spread function (PSF).


Without the necessary point sources to allow for a robust alignment between epochs we return to the method employed in our previous work (\cite{Williams2018}) and measure the diameter of the remnant such that perfect alignment is unnecessary. 

\section{Measurements and Discussion}
\label{measurements}
\begin{figure}
    \centering
    \includegraphics[width=\columnwidth]{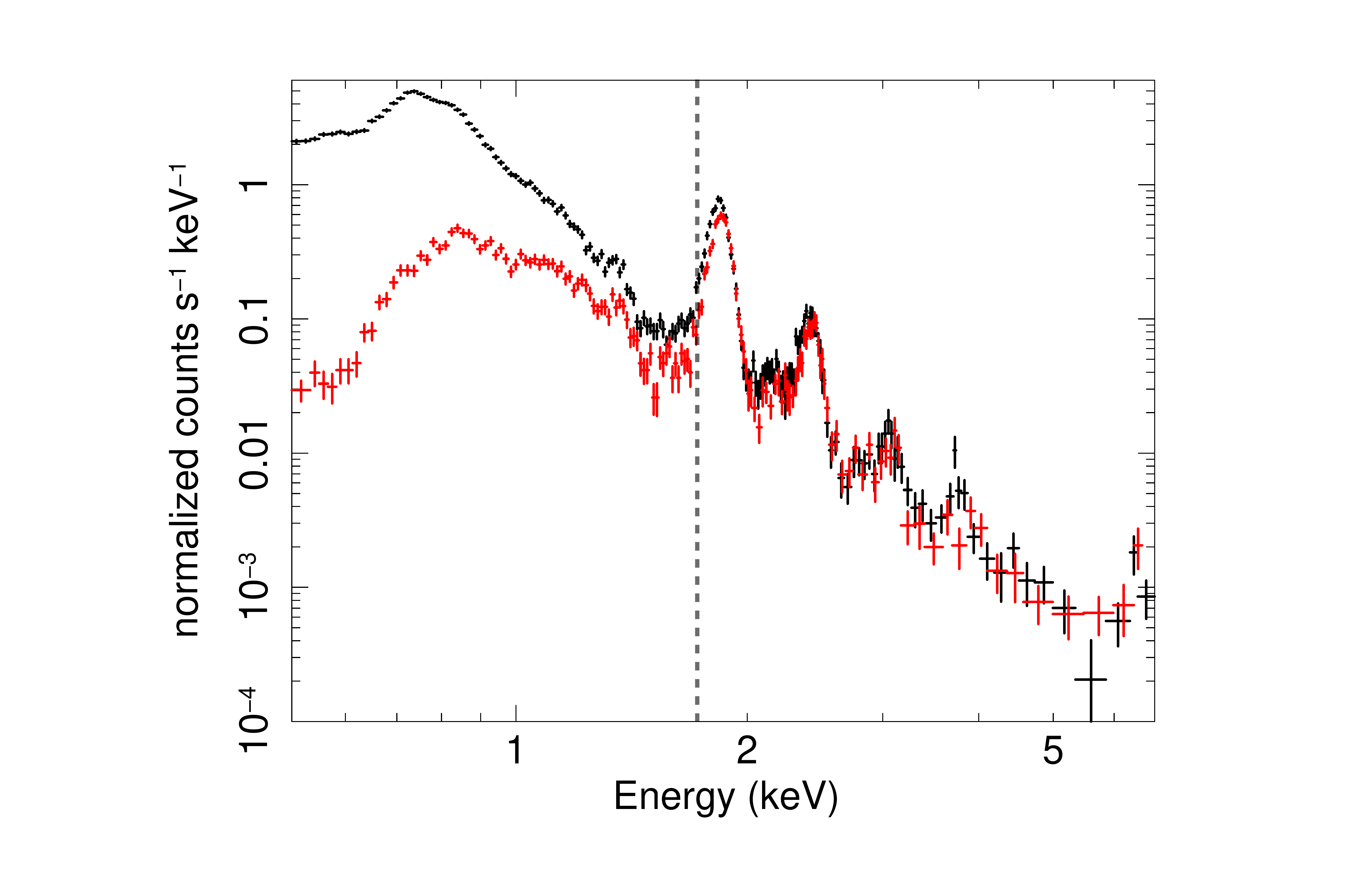}
    \caption{Spectrum taken from the 2000 observation (black) and the longest individual 2020 observation ObsID 24637 (red) showing the decrease in low energy sensitivity. The dotted line is at 1.7 keV, above which the observations may be compared fairly.)}
    \label{fig:Spectrum}
\end{figure}

The low energy sensitivity of \textit{Chandra} has been significantly affected by a build-up of contaminants (\cite{Marshall2004}). In order to achieve a fair comparison between epochs we filter the observations to energies between 1.7 and 7 keV (Figure \ref{fig:Spectrum}).

\begin{figure}
    \centering
    \includegraphics[width=\columnwidth]{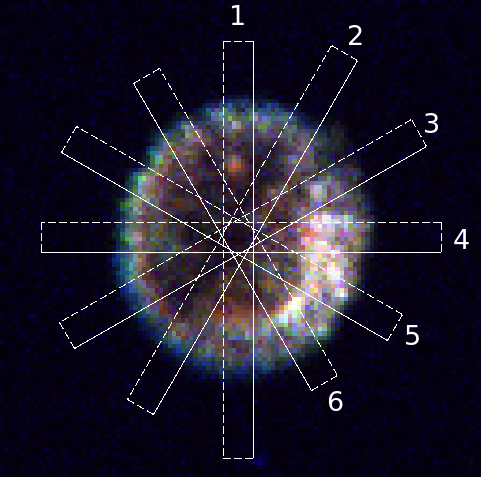}
    \caption{Chandra X-ray image of 0509-67.5, with 0.5-1.2 keV in red, 1.2-2 keV in green, and 2-10 keV in blue, overlaid with our 6 profile extraction regions as described in the text. Each region is 7 pixels wide where 1 pixel is the native Chandra pixel scale of 0$^{\prime\prime}$.492. The region numbers are positioned at the right side of the profiles in Figure \ref{fig:SampleProjections}.}
    \label{fig:Projections}
\end{figure}

We measured the diameter of the remnant using radial profiles along 6 regions which all cross the geometric center of the remnant and are spaced by rotations of 30$^o$ (Figure \ref{fig:Projections}). Each region is 7 pixels wide, where 1 pixel is the native Chandra pixel size of 0$^{\prime\prime}$.492. The measurements were made individually for the 6 longest 2020 observation segments (Table \ref{tab:obs}) each paired with the 2000 observation.

\begin{figure}
    \centering
    \includegraphics[width=\textwidth]{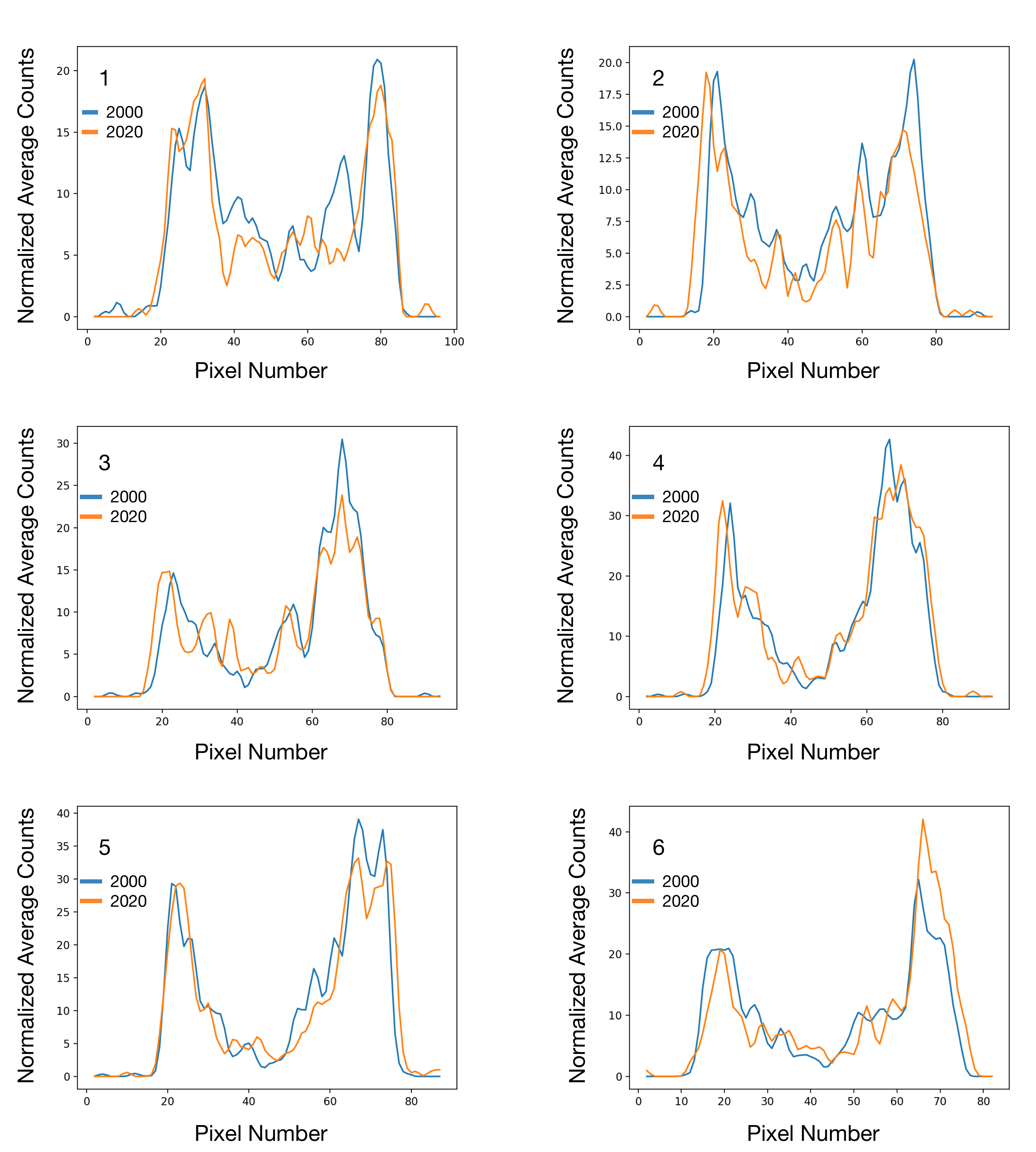}
    \caption{Sample projections from each region ordered clockwise starting with the north-south projection as numbered in Figure \ref{fig:Projections}. The data are independently normalized prior to fitting the shift required for each side. The projections displayed are normalized for the left side. The right side of the profiles are the ends of the regions closest to the identifying region numbers shown in Figure \ref{fig:Projections}.}
    \label{fig:SampleProjections}
\end{figure}

    The forward shock front is assumed to be the location in the profile where the brightness rises sharply. Profiles are normalized to align the tops of these rise peaks (Figure \ref{fig:SampleProjections}). We then shift epoch 1 with respect to epoch 2 on a grid of 0$^{\prime\prime}$.0048 resolution elements corresponding to ~0.01 \textit{Chandra} pixels and minimize the chi-squared value resulting from the difference between the two profiles over small few pixel windows.  The best fit shift depends on the choice of fitting window, which is impossible to state with absolute certainty. To account for this uncertainty, for each measurement several window values were chosen spanning between roughly 4-10 pixels depending on the specific profile shape, beginning where the profile rises from the background and ending at the outermost peak. For each individual measurement, we change the window by 1 pixel on each end and average the results from repeating this process 3 times. For each projection angle we then have a set of required shifts from each of the 6 longest 2020 observations paired with the 2000 observation. The diameter expansion is then determined by the difference between the required shifts for each pair of shock fronts on opposite sides of the remnant. In this method, the absolute position of either shock is irrelevant, and our results are independent from the choice of alignment. We report the average radial expansion among the 6 pairs of 2000 - 2020 observations and the associated standard deviation and expansion velocity in km s$^{-1}$ in Table \ref{tab:ExpansionVelocity} using a distance of 50 kpc to the LMC (\cite{Pietrzy2013}). The standard deviation column is calculated from the expansion results from each of the 6 observation pairs. The velocity we measure results from the expansion of the diameter of the remnant. Since this expansion may be happening asymmetrically, reporting an average velocity with an error calculated by assuming all of the measurements are corresponding to the same true inherent value is not correct. Here we choose to report the average velocity with an error calculated from the deviation of the individual velocity measurements.

\begin{table}[]
    \centering
    \begin{tabular}{c c c c c}
        \hline
        Region & Expansion (pixels) & Standard deviation (pixels) & Expansion Velocity (km s$^{-1}$) & Standard deviation (km s$^{-1}$) \\\hline
        1	&	1.47	&	0.32	&	4350 & 960	\\
2	&	2.60	&	0.82	&	7710	& 2440 \\
3	&	1.88	&	0.66	&	5580	& 1970 \\
4	&	2.31	&	0.54	&	6840	& 1590 \\
5	&	2.32	&	0.47	&	6880	& 1390 \\
6	&	1.81	&	0.69	&	5380	& 2060 \\\hline
\textbf{Avg}	&	2.07	&		&	\textbf{6120} & \textbf{1200}	\\
        \hline
    \end{tabular}
    \caption{Expansion measurements for the 6 individual diameter projections and resulting average. The 6 velocities are not considered the true velocity since the remnant may be expanding asymmetrically. We therefore consider the deviation of the individual velocity results as a method to estimate our errors.}
    \label{tab:ExpansionVelocity}
\end{table}


As a sanity check, we calculated the average expansion from the area of contours for each epoch. This is a completely independent method. To examine the effect of the differing response of the detector particularly at low energies, we first equate the number of detected 0.5-7 keV counts in the 2000 observation counts image and the longest 2020 observation segment by filtering the 2000 observation exposure time to a brief period which resulted in an effective exposure time of 6.9 ks. We then drew a single contour in DS9 around the remnant defining a contour level of 0.25 with a smoothing level of 4. The contours were then converted to region files from which the areas were read. The 2000 contour encompassed 704.2 square arcseconds while the 2020 contour covered 771.0. The area increases as the square of the diameter, meaning that on average, the diameter of the remnant increased by 1.4$^{\prime\prime}$ or 2.8 pixels. This is larger than the diameter increase we measure with the projection method. The brief effective exposure time however leads to a systematically smaller area since the leading edge of the remnant appears to be dominated by harder emission which would not have been able to build up to the same degree as the longer 2nd epoch. We then restricted the energy range to the 1.7-7 keV band where the change in response is much less significant. We again filter the exposure time of the 2000 observation in order to match the total number of counts in the longest 2020 observation segment. This both accounts for the small change in response visible in the differing heights of the Si line in Figure \ref{fig:Spectrum} and the different exposure times which affect the number of background counts to which the contour method is sensitive. Here we required an effective exposure of 31ks. The resulting contour areas increased from 713.1 square arcseconds as measured in the 2000 epoch to 757.7 square arcseconds in 2020. This corresponds to a diameter expansion of 1.9 pixels and is consistent with the result we found using the projection method, supporting our result.

\begin{figure}
    \centering
    \includegraphics[width=0.5\textwidth]{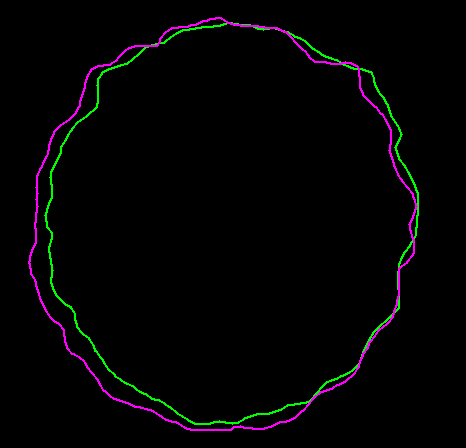}
    \caption{Contours containing the full remnant with a level of 0.25 created from the filtered 1.7-7 keV event files. The green contour is from the 2000 observation while the longest 2020 observation is magenta. The 2000 observation was filtered to a shorter exposure time to equate the number of counts with the 2020 segment. The contours use the native coordinates for each epoch and the offset highlights the difficulty with alignment.}
    \label{fig:Contours}
\end{figure}

\begin{figure}
    \centering
    \includegraphics[width=0.5\textwidth]{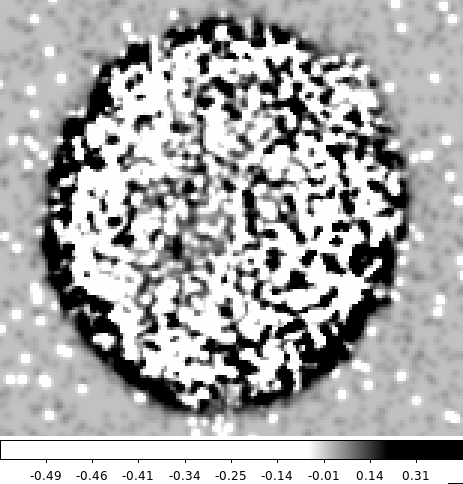}
    \caption{Difference image showing the aligned and merged 2020 observations minus the 2000 observation. Each image was filtered to 1.7-7 keV, smoothed with a 2 pixel gaussian, and the merged image was normalized to the number of counts in the filtered 2000 observation. Expansion is clearly visible, but the true alignment between epochs is not trusted.}
    \label{fig:DifferenceImage}
\end{figure}

As an additional test, we attempt to align the 2020 observations using the method of \cite{Carlton2011}. We smooth one counts image constructed from a single observation segment filtered to 0.4-8 keV, and fit the unsmoothed images from the other observations binned to 1/4 of an acis pixel to this smoothed image using only translations and no expansion. We merged the aligned 2020 observations using the CIAO tool merge\_obs and repeated our analysis using the diameter measurement method. We find results consistent with our unaligned individual measurements indicating that we have achieved a reasonable alignment of the 2020 observations. We attempted to align with the 2000 observation using the added counts in the few point sources yielded by the merging. A resulting difference image is shown in Figure \ref{fig:DifferenceImage}. While expansion is clearly visible in the difference image, there also appears to be a systematic shift between epochs suggesting the alignment is not sufficient to perform localised velocity measurements.

The high shock velocity we find supports SNR 0509--67.5's status as a young SNR. The average velocity we calculate corresponds to an undecelerated free expansion age of 600 years with the true age somewhat lower. The velocity we measure is consistent with the results of \cite{Hovey2015} who then used hydrodynamic simulations to calculate an age of 310 years. Our method, which measures the change in the diameter of the remnant, does not allow for detailed inferences on whether one area of the remnant is expanding more quickly than another. The difference in brightness of the south western limb of the remnant may suggest a local increase in density in this direction. However, our diameter measurement normal to this direction does not yield systematically different results than the others. This is not unexpected since the measurement includes the motion in the opposite direction from the opposing shock which would offset any difference in velocity.
We list the velocity and age we infer along with the X-ray proper motion results from comparable Type Ia remnants in Table \ref{tab:VelocityComp}. We find a comparable velocity to Tycho's remnant and the non-thermal rims of SN 1006, both of which are expanding into low density media similar to SNR 0509-67.5.

\begin{table}[]
    \centering
    \begin{tabular}{l c c}
    \hline
        Remnant &  Velocity (km s$^{-1}$) & Age (yr)\\\hline
        0509-67.5 & 6120   & 620$*$ \\
        N103B$^{a}$ &  4170 & 850$*$ \\
        Tycho$^{b}$ & 5300  & 450 \\
        Kepler$^{c}$ & 1780  & 418 \\
        SN 1006$^{d}$ & 3000 - 5000 & 1016 \\\hline
        $*$ Free expansion \\
        $^{a}$ \cite{Williams2018} & & \\
        $^{b}$ \cite{Williams2016} & & \\
        $^{c}$ \cite{Coffin2022} & & \\
        $^{d}$ \cite{Katsuda2013,Winkler2014} & & \\\hline
    \end{tabular}
    \caption{Comparison of SNR 0509-67.5 with the X-ray proper motion expansion velocity measurements of other young Type Ia remnants. The ages inferred from the measured velocity assuming free expansion represent the upper limit on the age of SNRs 0509-67.5 and N103B, while the age of Tycho, Kepler, and SN 1006 are from historical record (e.g. \cite{Baade1943,Brown1952,Gardner1965}).}
    \label{tab:VelocityComp}
\end{table}

\section{Conclusions}
We have observed the LMC SNR 0509--67.5, 20 years after its initial Chandra observation with the goal of measuring the expansion of the remnant. The lack of strong point sources in the field made the absolute astrometric alignment impossible. We measure instead the change in the diameter of the remnant in six directions, yielding an average expansion velocity of just over 6100 km s$^{-1}$. This corresponds to an undecelerated age of 600 yrs, making the true age even lower. Our results are broadly consistent with previous {\it Chandra} X-ray measurements, as well as those made in H$\alpha$ with {\it HST}. Continued monitoring of young SNRs in the Galaxy and the LMC is critical for understanding the early stage of their development, the stage most closely related to the explosion of the progenitor system. We encourage future observations of this remnant and others like it with {\it HST} and {\it LUVOIR} in optical bands and {\it Chandra}, {\it AXIS}, and {\it Lynx} at X-ray wavelengths.

We acknowledge support from the Smithsonian Astrophysical Observatory / NASA grants GO0-21057 and GO0-21057B. B.G acknowledges the material is based upon work supported by NASA under award number 80GSFC21M0002. We thank the referee for their careful reading of the paper which improved its quality and clarity.

%

\vspace{5mm}
\facilities{CXO}







\bibliography{References.bib}
\bibliographystyle{aasjournal}



\end{document}